\documentclass[superscriptaddress, reprint,
nofootinbib,
bibnotes,
 amsmath,amssymb,
aps,
]{revtex4-1}

\usepackage{graphicx}% Include figure files
\usepackage{dcolumn}% Align table columns on decimal point
\usepackage{bm}% bold math
\usepackage{hyperref}% add hypertext capabilities
\usepackage[mathlines]{lineno}% Enable numbering of text and display math
\usepackage{xcolor}

\usepackage{mathtools}

\newcommand{\ghost}[1]{}

\begin{document}

\title{When the Environment Speaks: Quantum Signatures in Non-Attractor Inflation}% Force line breaks with \\

\author{Mattia Cielo}
 \email{mattia.cielo@su.se}
\affiliation{Nordita AlbaNova Univ. Center, Hannes Alfvéns väg 12, SE-106 91 Stockholm, Sweden}%

\author{Simone Scarlatella}
 \email{simone.scarlatella@unina.it}
\affiliation{%
 Dipartimento di Fisica “Ettore Pancini”, Università degli studi di Napoli “Federico II”, Complesso Univ. di Monte S. Angelo, I-80126 Napoli, Italy
}
 \affiliation{INFN - Sezione di Napoli, Complesso Univ. di Monte S. Angelo, I-80126 Napoli, Italy}%Lines break automatically or can be forced with \\

\author{Gianpiero Mangano}%
 \email{gianpiero.mangano@na.infn.it}
\affiliation{%
 Dipartimento di Fisica “Ettore Pancini”, Università degli studi di Napoli “Federico II”, Complesso Univ. di Monte S. Angelo, I-80126 Napoli, Italy
}
 \affiliation{INFN - Sezione di Napoli, Complesso Univ. di Monte S. Angelo, I-80126 Napoli, Italy}%Lines break automatically or can be forced with \\

\author{Ofelia Pisanti}
 \email{ofelia.pisanti@na.infn.it}
\affiliation{%
 Dipartimento di Fisica “Ettore Pancini”, Università degli studi di Napoli “Federico II”, Complesso Univ. di Monte S. Angelo, I-80126 Napoli, Italy
}
 \affiliation{INFN - Sezione di Napoli, Complesso Univ. di Monte S. Angelo, I-80126 Napoli, Italy}%Lines break automatically or can be forced with \\

\date{\today}% It is always \today, today,

\begin{abstract}
We study the open quantum dynamics of the adiabatic curvature perturbation interacting 
with a massive entropic scalar environment during an inflationary scenario featuring a 
transient Ultra-Slow-Roll phase. Working within a Gaussian two-field effective 
Lagrangian, we employ the exact Transport Equations Method to track the full 
non-unitary, non-Markovian evolution of the system's covariance matrix across the 
SR-USR-SR transition. We find that the efficiency of decoherence is 
sensitive to the background kinematics at horizon crossing.
Most importantly, the interaction with the environment leaves distinct observable 
imprints on the primordial scalar power spectrum: the characteristic interference 
dip preceding the USR-driven enhancement can be partially or completely erased, the 
growth slope modified, and oscillatory features induced near the peak. 
Propagated to second order, these distortions further imprint on the stochastic 
background of Scalar-Induced Gravitational Waves, breaking single-field predictions 
and yielding unique spectral signatures potentially accessible to LISA. Our results 
demonstrate that the quantum environment is not a passive spectator during inflation, 
but an active agent whose imprint on the primordial universe may be within reach of 
the next generation of cosmological observations.

\end{abstract}

\maketitle

%\tableofcontents

\section{Introduction}
\label{sec:intro}

Cosmological inflation not only solves the horizon and flatness problems of the standard Big Bang model but also provides a compelling causal mechanism for the origin of large-scale structures via the amplification of vacuum quantum fluctuations \cite{PhysRevD.23.347, Starobinsky:1980te, Linde:1981mu, Liddle_1994, osti_6268106, Allen:1985ux, MUKHANOV1992203, Cheung:2007st, Lemoine:2008zz, Wang:2013zva, Baumann_2022}. Recently, there has been a surge of interest in inflationary models capable of generating a significant enhancement in the primordial power spectrum at small scales. Such an amplification is a fundamental prerequisite for the formation of Primordial Black Holes (PBHs) upon horizon reentry during the post-inflationary radiation-dominated era, making them highly attractive candidates for cold dark matter~\cite{Carr_2016, Carr_2020, Carr_2021, Pi:2022zxs, GARCIABELLIDO201747, Garcia-Bellido:1996mdl,  Pattison:2017mbe, Clesse:2017bsw, Ezquiaga_2018, Bartolo:2018evs,LISACosmologyWorkingGroup:2023njw, Ozsoy:2023ryl, Ballesteros:2017fsr}. 

The standard mechanism to achieve this localized power enhancement relies on a transient phase of Ultra-Slow-Roll (USR) inflation, sandwiched between two standard Slow-Roll (SR) regimes. During the USR phase, the inflaton traverses an extremely flat region of its potential, slowing down considerably due to Hubble friction and leading to a macroscopic growth of the curvature perturbations ~\cite{Leach:2001zf, Kinney:2005vj,Byrnes:2018txb, Raveendran_2022, Cielo:2024poz}.

While the classical dynamics and the resulting primordial power spectrum in USR models have been extensively studied, the fundamentally quantum nature of these perturbations requires a deeper quantum information-theoretic approach. The cosmological perturbations are born as pure quantum vacuum states, yet cosmological observations strongly suggest a classical, stochastic distribution of structures. Understanding this quantum-to-classical transition, often referred to as decoherence, is a long-standing challenge \cite{Brandenberger:1990bx, Polarski:1995jg, Kiefer:1998qe, Burgess:2006jn, Martin_2018_a, Martin_2018_b, Martin:2021znx, Burgess:2022nwu, Colas:2022kfu, Colas_2024a,Kranas:2025jgm, lopez2025quantumsignaturesdecoherenceinflation, Brahma_2022_a, Brahma:2025dio, Brahma:2026qtm, Micheli:2022tld, Martin:2021znx, Martin_2018_a, Blencowe:2012mp, Christie:2025knc, Brandenberger:1990bx, Colas_2024a}. A rigorous way to tackle this is within the framework of Open Quantum Systems \cite{Hollowood:2017bil,Colas:2022hlq, Kaplanek:2025moq, Brahma_2022_b}. In realistic multi-field scenarios, the observable curvature (adiabatic) perturbations inevitably interact with unobservable heavy degrees of freedom, such as isocurvature/entropic modes. Tracing out these unobserved environmental sectors induces decoherence and entropy production in the reduced state of the system \cite{Assassi:2013gxa, Brahma:2024ycc, Colas_2024a, Colas:2022hlq, Colas:2022kfu, Boyanovsky_2018, Hollowood:2017bil, Boyanovsky_2015, Christie:2025knc, Martin_2018_b, Martin_2018_a, Burgess:2024eng, Burgess:2022nwu, Haque:2026bby, cielo2025quantumrecoherencepresenceexcited}.

In this work, we investigate the quantum information properties of the primordial power spectrum in an inflation model featuring a SR-USR-SR transition. We consider an effective two-field Lagrangian where the adiabatic perturbation couples to a massive entropic environment. By numerically solving the transport equations for the covariance matrix of the Gaussian state, we accurately track the full non-Markovian dynamics of the system. We compute the primordial power spectrum and extract the exact evolution of the quantum purity and von Neumann entanglement entropy, exploring how the environmental mass $m$ and the interaction coupling $\lambda$ affect the decoherence and recoherence processes of the modes as they cross the horizon during the USR phase. 

Recently, the dynamics of quantum decoherence and recoherence in the early universe has gathered significant attention in the literature (see, e.g.\cite{Colas:2022kfu, Brahma:2024ycc}). Building upon these foundational efforts, our study represents a systematic investigation of how the loss of coherence and the evolution of entropy strictly depend on the precise moment the modes exit the Hubble radius relative to the SR-USR-SR transitions. Furthermore, unlike previous analyses that focused solely on the quantum state of the modes \cite{Brahma:2024ycc}, we explicitly compute the resulting primordial power spectrum. We demonstrate that the interaction with the quantum environmental fields is not merely a formal feature, but it imprints distinctive observable deviations and modifications on the spectrum's shape and amplitude. Finally, we propagate these environment-induced features to late-time observables by calculating the stochastic background of scalar-induced gravitational waves (SIGWs). We show that the quantum nature of the entropic sector during the USR phase can leave significant phenomenological signatures on the GW spectrum, potentially falling within the sensitivity range of future space-based interferometers \cite{LISACosmologyWorkingGroup:2023njw, LISACosmologyWorkingGroup:2022jok, LiteBIRD:2022cnt, Campeti:2020xwn, Guzzetti:2016mkm, Micheli:2022tld, Ghaleb:2025xqn, LISACosmologyWorkingGroup:2025vdz, Marriott-Best:2024anh}.
The rest of the paper is organized as follows. In Sec.~\ref{sec:model}, we outline the background USR dynamics and the two-field effective Lagrangian, introducing the transport equations formalism used to evolve the quantum perturbations. In Sec.~\ref{sec:quantuminfo}, we define the relevant quantum information measures, followed by Sec.~\ref{sec:spectra} where we present our numerical results for the primordial power spectra and the resulting scalar-induced gravitational waves, before drawing our final conclusions in Sec.~\ref{sec:conclusions}.

\textit{Notation.} Throughout this paper, we work in $(3+1)$-dimensions with a $(-,+,+,+)$ metric signature. We employ natural units $\hbar = c = 1$ and set the reduced Planck mass $M_{\rm Pl} = (8\pi G)^{-1/2} = 1$. The amount of cosmic expansion is measured in terms of the number of e-folds $N = \ln(a/a_0)$, where $a$ is the scale factor and $a_0$ its initial value.

\section{The Model}
\label{sec:model}

\subsection{Background Dynamics}
\label{sec:background_usr}

Assuming homogeneity and isotropy on sufficiently large scales, the background spacetime is described by the spatially flat Friedmann-Robertson-Walker (FRW) metric,
\begin{equation}
    ds^2 = -dt^2 + a^2(t)\delta_{ij}dx^i dx^j \, ,
\end{equation}
where $t$ is the cosmic time. The expansion rate of the universe is characterized by the Hubble parameter $H \equiv \dot{a}/a$. The background dynamics is driven by a single canonical scalar field, the inflaton $\Phi(t)$, rolling down its potential $V(\Phi)$. The evolution of the system is governed by the Friedmann and Klein-Gordon equations\footnote{As usual, dot and prime denote derivatives with respect to cosmic time $t$ and conformal time $\tau$, respectively.}:
\begin{align}
    H^2 &= \frac{1}{3} \left( \frac{1}{2}\dot{\Phi}^2 + V(\Phi) \right) \, , \label{eq:friedmann} \\
    \ddot{\Phi} &+ 3H\dot{\Phi} + V_{,\Phi} = 0 \, , \label{eq:KG}
\end{align}
where $V_{,\Phi} \equiv dV/d\Phi$. To achieve the required 60 or so e-folds of inflation~\cite{Baumann_2022}, standard slow-roll conditions require the potential energy to dominate over the kinetic energy ($\dot{\Phi}^2 \ll V$) and the field acceleration to be negligible ($\ddot{\Phi} \ll 3 H \dot{\Phi}$). This regime is usually described through the hierarchy of Hubble flow functions, \cite{Dimopoulos:2017ged, CHENG2022136956, Kinney:2005vj, Tasinato:2023ukp}:
\begin{equation}
    \epsilon_0 \equiv -\frac{H'(N)}{H(N)} \, , \qquad\epsilon_{n+1} \equiv \frac{d \ln |\epsilon_n|}{dN} \, , \quad n \geq 0 \, .
\end{equation}
For $n=1$, one finds the first SR parameter $\epsilon_1 \equiv \epsilon = -\dot{H}/H^2$, and for $n=2$ the second SR parameter $\eta \equiv \dot{\epsilon}/(H \epsilon)$. Inflation takes place provided $\epsilon < 1$, and standard SR inflation is defined by the condition $|\epsilon_n| \ll 1$ for all $n > 0$.

While SR inflation generates a nearly scale-invariant power spectrum, sourcing Primordial Black Holes (PBHs) \cite{Carr_2016, Carr_2020, Carr_2021, GARCIABELLIDO201747, Garcia-Bellido:1996mdl, Ezquiaga_2018, Pattison:2017mbe} requires a strong amplification of the perturbations at specific scales (typically reaching a power spectrum $\sim \mathcal{O}(10^{-2})$). This is achieved by transiently violating the SR conditions through an USR phase. 

\begin{figure}[h!]
\centering
\includegraphics[width=\linewidth]{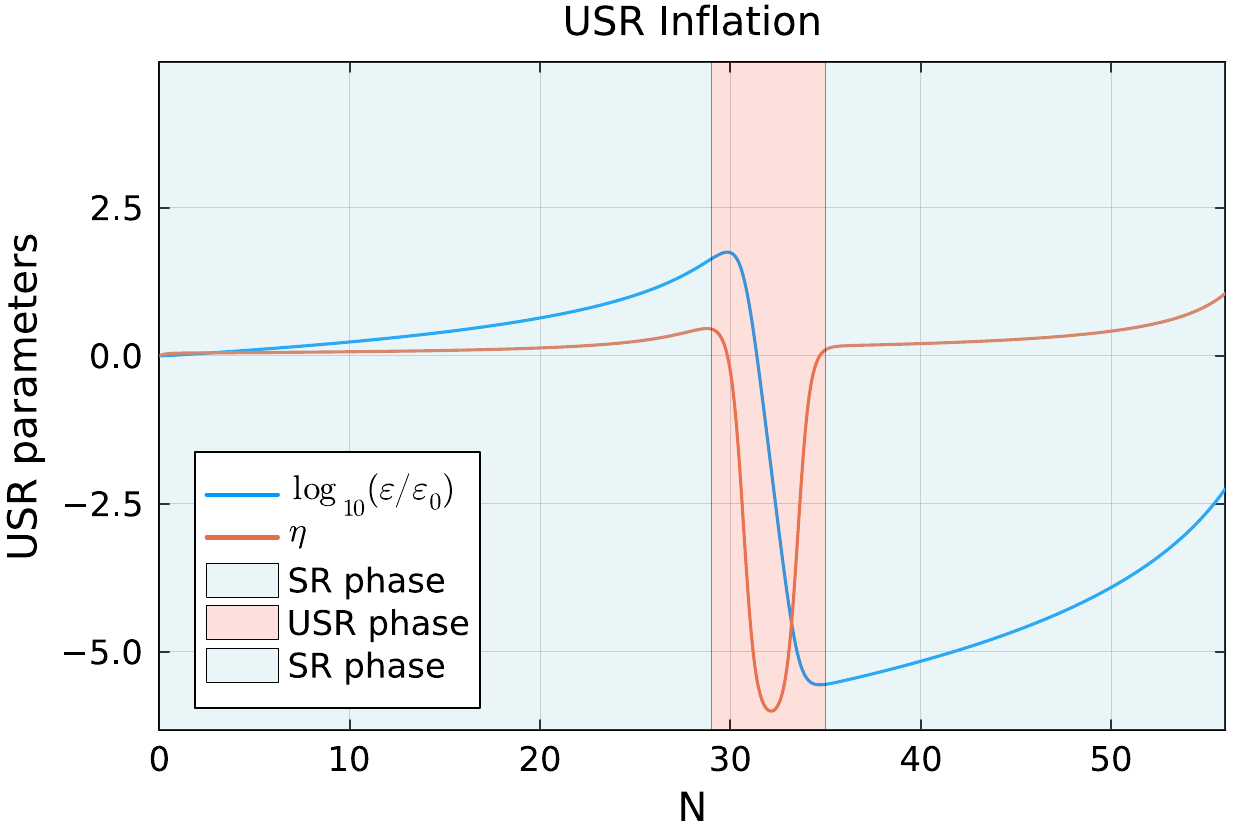}
\caption{Evolution of the SR parameters showing the sharp SR-USR-SR transition. The second SR parameter $\eta$ reaches $-6$ during the USR phase, driven by Hubble friction.}
\label{fig:usr-background}
\end{figure}

During the USR regime, the inflaton dynamics is no longer driven by the potential, which becomes extremely flat locally ($V_{,\Phi} \simeq 0$). Instead, the dominant term in Eq.~\eqref{eq:KG} becomes the Hubble friction $3 H \dot{\Phi}$. Consequently, the second SR parameter strongly violates the SR condition ($|\eta| \sim \mathcal{O}(1)$), specifically dropping to a value of $\eta \simeq -6$, while $\epsilon$ exponentially decreases. This sharp SR-USR transition, followed by a return to SR where $\eta$ becomes small again, causes a macroscopic growth in the curvature perturbations. The typical evolution of these background parameters across the different phases is shown in Fig.~\ref{fig:usr-background}.

A single-field scalar potential with a near-inflection point naturally accommodates this sequence. These potentials arise frequently in models like Higgs Inflation~\cite{Bezrukov_2008, Garc_a_Bellido_2009} and take advantage of asymptotic flatness for large field values, in agreement with CMB observations. In the present work, we adopt the following functional form to generate the necessary spike in the power spectrum~\cite{Garcia-Bellido:2017mdw, Ezquiaga_2018, CHENG2022136956}:
\begin{equation}
    V(\Phi) = \frac{\lambda}{12} \Phi^2 \nu^2 \frac{6 - 4a\frac{\Phi}{\nu} + 3\left(\frac{\Phi}{\nu}\right)^2}{\left(1 + b\left(\frac{\Phi}{\nu}\right)^2\right)^2} \, ,
    \label{eq:USR_potential}
\end{equation}
with the parameter choice $\lambda = 1.86 \cdot 10^{-6}$, $\nu = 0.196$, $a = 0.7071$, and $b = 1.5$. We solve the exact background equations~\eqref{eq:friedmann} and~\eqref{eq:KG} using the number of e-folds $N$ as the independent variable. The resulting background functions $a(N)$ and $H(N)$ accurately capture the kinematics across the transition and serve as input for the transport equations governing the quantum perturbations.

\subsection{\label{subsec:model}Two-field Effective Lagrangian}

Current cosmological observations, in particular from the Cosmic Microwave Background (CMB) \cite{Planck:2018vyg, BICEP:2021xfz,calabrese2025atacamacosmologytelescopedr6}, strongly indicate that the primordial power spectrum is predominantly sourced by adiabatic fluctuations, while isocurvature modes are highly suppressed. However, relying solely on the adiabatic sector leaves a critical conceptual gap regarding the classicalization of these initial quantum vacuum fluctuations. 

A rigorous framework to describe this quantum-to-classical transition, often referred to as decoherence, treats the inflationary perturbations as an open quantum system. In this scheme, the observable adiabatic perturbation (the \textit{system} S) couples at leading order to a hidden entropic field (the \textit{environment} E) evolving on a (quasi-)de Sitter background \cite{Starobinsky:1980te,Starobinsky:1992ts}. Recently, open effective field theory (EFT) approaches have been successfully adopted to study diffusion and entanglement for non-attractor USR backgrounds (see, e.g. \cite{Brahma:2024ycc, Brahma:2024yor}), making this a highly suitable framework for our analysis.

The effective Lagrangian density considered to model this interaction is \cite{Assassi:2013gxa}:
\begin{eqnarray}
\mathcal{L} &=& 
a^{2}\epsilon M_{\rm Pl}^{2}(\zeta')^{2} 
- a^{2}\epsilon M_{\rm Pl}^{2}(\partial_{i}\zeta)^{2} \nonumber \\
&+& \tfrac{1}{2}a^{2}(\mathcal{F}')^{2} 
- \tfrac{1}{2}a^{2}(\partial_{i}\mathcal{F})^{2} 
- \tfrac{1}{2}\mu^{2}a^{4}\mathcal{F}^{2} \nonumber \\
&+& \lambda a^{3}\sqrt{2\epsilon}\,M_{\rm Pl}\,\zeta' \mathcal{F} \, ,
\label{eq:L-zeta-F}
\end{eqnarray}
where $\epsilon \equiv -\dot H/H^{2}$ is the first slow-oll parameter, $M_{\rm Pl}$ is the reduced Planck mass, $\mu$ the bare mass of the environment field, and $\lambda$ is a dimensionful interaction parameter. The $\zeta$ field represents the adiabatic perturbation (the system), while the $\mathcal{F}$ field is the environment with which the system interacts, mediating the decoherence process.

It is convenient to introduce the canonical Mukhanov-Sasaki fields by scaling both quantities to obtain Sasaki-like equations:
\begin{equation}
z^{2} \;\equiv\; 2 a^{2} \epsilon M_{\rm Pl}^{2}\,, 
\qquad v \;\equiv\; z\,\zeta\,,
\qquad u \;\equiv\; a\,\mathcal{F}\,,
\label{eq:canon-defs}
\end{equation}
for which the free parts of Eq.~\eqref{eq:L-zeta-F} take their standard canonical form, and the background ``squeezing'' terms are encoded in $z''/z$ and $a''/a$. Using $\zeta'=(v'/z) - (z'/z^{2})v$ and $\mathcal{F}=u/a$, the mixing term becomes
\begin{equation}
\mathcal{L}_{\rm mix} \;=\; \lambda\,a\;\Big(v' - \frac{z'}{z}\,v\Big)\,u\,.
\label{eq:Lmix-vu}
\end{equation}

Since the mixing is linear in $v'$ and $v$, the theory remains globally Gaussian and amenable to exact mode-by-mode analyses. The conjugate momenta are
\begin{equation}
\pi_S \;\equiv\; \frac{\partial \mathcal{L}}{\partial v'} \;=\; v' -\frac{z'}{z}v+ \lambda\,a\,u\,,
\quad
\pi_E \;\equiv\; \frac{\partial \mathcal{L}}{\partial u'} \;=\; u' - \frac{a'}{a}u\,,
\label{eq:conj-momenta}
\end{equation}
where the shift in $\pi_S$ is induced strictly by the derivative mixing in Eq.~\eqref{eq:Lmix-vu}. The Hamiltonian density, $\mathcal{H} \equiv \pi_S v' + \pi_E u' - \mathcal{L}$, can therefore be written as
\begin{eqnarray}
\mathcal{H} &=& \tfrac{1}{2}\!\Big[\pi_S^{2} + (\partial_i v)^{2} + \tfrac{z'}{z}\{\pi_S,v\}\Big] \nonumber \\
&+& \tfrac{1}{2}\!\Big[\pi_E^{2} + (\partial_i u)^{2} + (\mu^{2}+\lambda^{2})a^{2}u^{2} + \tfrac{a'}{a}\{\pi_E,u\}\Big] \nonumber \\
&-& \lambda a\,\pi_S u \, ,
\label{eq:H-density}
\end{eqnarray}
which is quadratic and mode-separable in Fourier space. This form makes transparent both the shifted kinetic structure induced by the mixing and the background-induced effective mass. 

In order to quantize the theory, we use the interaction picture so that the equations of motion for the two rescaled fields are two uncoupled Mukhanov-Sasaki differential equations:
\begin{subequations}
\begin{eqnarray}
v_{k}'' + \Big(k^{2} - \frac{z''}{z}\Big)\,v_{k}
&\;=\; 0\,,
\label{eq:eom-vk}\\[4pt]
u_{k}'' + \Big(k^{2} + m^{2} a^{2} - \frac{a''}{a} \Big)\,u_{k}
&\;=\; 0\,,
\label{eq:eom-uk}
\end{eqnarray}
\end{subequations}
where $m^2=\mu^2+\lambda^2$ is the effective mass \cite{MUKHANOV1992203, Mukhanov:2007zz}.

Assuming pure de Sitter kinematics deep inside the horizon to select the initial vacuum state, these two equations admit the following expressions as mode functions:
\begin{subequations}
\begin{eqnarray}
f_{k}^{(S)}(\tau) &=& \frac12 e^{i \frac{\pi}{2} (\nu_S + \frac12)} \sqrt{-\pi \tau} \, H^{(1)}_{\nu_S}(-k\tau \,) \,, 
\label{eq:eom-solvk}\\[4pt]
f_{k}^{(E)}(\tau) &=& \frac12 e^{i \frac{\pi}{2} (\nu_E + \frac12)} \sqrt{-\pi \tau} \, H^{(1)}_{\nu_E}(-k\tau \,) \,,
\label{eq:eom-soluk}
\end{eqnarray}
\end{subequations}
where $\nu_S^2=\frac94$ and $\nu_E^2=\frac94 - \frac{m^2}{H^2}$.
Although in a general inflationary background the index $\nu$ is sourced by a hierarchy of Hubble-flow parameters\footnote{Rewriting the mode equation in the standard Bessel form $v_k'' + \big(k^2 - (\nu^2-1/4)/\tau^2\big)v_k = 0$, and using the quasi-de Sitter relation $(a'/a)^2 \simeq 1/\tau^2$, the exact expression for $z''/z$ translates directly into the index $\nu^2 = \frac{9}{4} - \epsilon_1 + \frac{3}{2}\epsilon_2 + \frac{1}{4}\epsilon_2^2 - \frac{1}{2}\epsilon_1\epsilon_2 + \frac{1}{2}\epsilon_2\epsilon_3$ , see \cite{Maggiore:2018sht, Baumann_2022, Peter:2013avv, Brahma:2026qtm}.}, giving rise to degeneracies known as Wands dualities \cite{Wands:1998yp}, this structure is not exploited here. We restrict ourselves to the leading, pure de Sitter expressions for $\nu_S$ and $\nu_E$, as these are strictly used to fix the Bunch--Davies initial conditions deep inside the horizon. This leading-order treatment does not compromise the accuracy of the subsequent evolution: the transport equations themselves are built using the exact expression for $z'/z$, which remains exact in any background while being significantly simpler and linear in the slow-roll parameters.

The Gaussianity of the total Hamiltonian~\eqref{eq:H-density} allows us to fully track the non-Markovian open dynamics of the system by tracing over the environment $E$, resolving the covariance matrix without relying on perturbative approximations.

\subsection{\label{subsec:transpeq}Evolution: Transport Method}

To solve the full dynamics encompassing both the system and the environment sectors, we employ the Transport Equations Method (TEM), \cite{Anderson:2012em, Pinol:2023oux, Brahma:2024ycc, Colas:2022kfu}.
The theory described in the previous section can be conveniently recast into matrix form. Defining a phase-space vector for the field variables, $q^T = (v, p_v, u, p_u)$, the total Hamiltonian matrix $H^{(\rm S+E)}$ can be decomposed into the free sectors ($H_{(S)}$ and $H_{(E)}$) and the interaction sector ($H_{\rm int}$):
\begin{equation}
H^{(\rm S+E)}(\tau) =
\begin{pmatrix}
H_{(S)}(\tau) & H_{\mathrm{int}}(\tau) \\[4pt]
H_{\mathrm{int}}^{T}(\tau) & H_{(E)}(\tau)
\end{pmatrix} \, ,
\label{eq:S5}
\end{equation}
where the sub-blocks are explicitly given by:
\begin{subequations}
\begin{align}
H_{(S)}(\tau) &=
\begin{pmatrix}
\dfrac{k^{2}}{a^{2}} & \dfrac{z'}{z} \\[4pt]
\dfrac{z'}{z} & 1
\end{pmatrix} \, ,
\\[4pt] 
H_{(E)}(\tau) &=
\begin{pmatrix}
k^{2} + (m^{2} + \lambda^{2})a^2 & \dfrac{a'}{a} \\[4pt]
\dfrac{a'}{a} & 1
\end{pmatrix} \, ,
\\[4pt] 
H_{\mathrm{int}}(\tau) &=
\begin{pmatrix}
0 & 0 \\[4pt]
-\lambda a & 0
\end{pmatrix} \, .
\label{eq:S6}
\end{align}
\end{subequations}

Within this formalism, the full dynamical information of the Gaussian state is captured by the covariance matrix. Denoting the reduced density matrix of the system as $\Tilde{\rho}(\tau) = \text{Tr}_{\mathrm{E}}\big(\rho_{\mathrm{S}}(\tau) \otimes \rho_{\mathrm{E}}(\tau_0)\big)$\footnote{Since there is no natural choice for the initial density matrices, we assume that the environment is stationary due to the separation of typical timescales (Born - Oppenheimer approximation), effectively acting as a reservoir that exchanges energy and information with the system.}, the covariance matrix reads:
\begin{equation}
    \Sigma^{(S)}_{ij}(\tau) \equiv \frac12\text{Tr}_{\rm S} \Big[ \{q_i(\tau),q_j^\dagger(\tau)\} \Tilde{\rho}(\tau_0) \Big] \, ,
    \label{eq:covmatrix}
\end{equation} 
with $i,j=1,2$. These elements encode all the physically relevant observables.

Rather than solving the master equation directly, it is computationally advantageous to solve the complete set of transport equations for the total unitary evolution:
\begin{equation}
    \frac{d}{d\tau}\Sigma^{(\rm S +E)} = \Omega H^{(\rm S +E)}\Sigma^{(\rm S +E)} - \Sigma^{(\rm S +E)}H^{(\rm S +E)} \Omega \, ,
    \label{eq:exacttm}
\end{equation}
where $\Omega$ is a block-diagonal symplectic matrix with antisymmetric entries $\bigl(\begin{smallmatrix} 0 & 1 \\ -1 & 0 \end{smallmatrix}\bigr)$. Note that the factor of 2 has been absorbed to correctly match the Heisenberg equations of motion.

In our numerical scheme, we solve this system using the number of e-folds $N$ as the independent time variable. We supplement the system \eqref{eq:exacttm} with an explicit differential equation for the determinant of the system's covariance matrix:
\begin{eqnarray}
    \frac{d}{dN}\text{det}\,\Sigma^{(\rm S)} &=& \Sigma_{11}^{(\rm S)} \frac{d \Sigma^{(\rm S)}_{22}}{dN} +\Sigma_{22}^{(\rm S)} \frac{d \Sigma^{(\rm S)}_{11}}{dN} \nonumber \\
    &-& 2\Sigma_{12}^{(\rm S)} \frac{d \Sigma^{(\rm S)}_{12}}{dN} \,.
\end{eqnarray}
While theoretically redundant, dynamically tracking $\text{det}\,\Sigma^{(\rm S)}$ circumvents severe numerical cancellations between diverging quantities at late times, making the computation of the quantum purity significantly more robust.

We initialize the system by imposing standard Bunch-Davies vacuum conditions on both the adiabatic and entropic covariance elements, matching the free Hankel mode solutions deep inside the horizon. This yields a coupled system of eleven differential equations (detailed in Appendix~\ref{app:sec:transpeq}). 

A key distinction of our methodology is the rigorous treatment of the background kinematics. While previous literature studying quantum markers often approximates the Hubble parameter $H$ as a constant (pure de Sitter), our analysis inherently requires capturing the transient dynamics of the USR phase. Therefore, we use the exact, time-evolving background functions $a(N)$ and $H(N)$ obtained from the background equations. This allows us to accurately evaluate the exact, mode-dependent resummed effects on the power spectrum and the quantum information markers across the SR-USR-SR transitions.

\section{\label{sec:quantuminfo}Quantum Information}

A useful way to characterize the quantum-to-classical transition of cosmological perturbations is through the purity and the entropic content of the reduced state, describing the adiabatic mode after tracing out the environment \cite{Eisert_2002, Boyanovsky_2018, Brahma_2022_a, Colas_2024a, Colas_2024b}. As already established, for a globally Gaussian state, the system's density matrix $\rho_S$ is completely determined by its covariance matrix $\Sigma^{(S)}_{ij}$. 

The \emph{purity} is defined as
\begin{equation}
    \gamma \equiv \mathrm{Tr}\big(\rho_S^2\big) = \frac{1}{4\,\det \Sigma^{(S)}} \, .
    \label{eq:purity}
\end{equation}
A pure state corresponds to $\gamma=1$, whereas $\gamma<1$ signals the emergence of decoherence. An intuitive way to understand how purity and decoherence are related is by assuming a Gaussian ansatz for the reduced system. Using the parametrization introduced in Ref.~\cite{Hollowood:2017bil}, the density matrix is expressed as:
\begin{equation}
    \rho_S=C \exp{ \Big( -\Omega(\tau)|v|^2-\Omega^*(\tau)|u|^2-\frac{\xi(\tau)}{2}|v-u|^2 \Big)}  \, ,
\end{equation}
with $\Omega =\Omega_R \, + i \, \Omega_I \in \mathbb{C}$ and $\xi \in \mathbb{R}$. It can be shown~\cite{Brahma:2024yor} that purity is directly tied to the decoherence parameter $\xi$:
\begin{equation}
    \gamma = \frac{1}{4\,\det \Sigma^{(S)}} = \Big( 1+ \frac{\xi}{\Omega_R} \Big)^{-1} \, .
\end{equation} 
This highlights how $\xi$, and in turn $\gamma$, quantifies the irreversible loss of quantum coherence. 

Another essential quantum informatic measure is the von Neumann entanglement entropy of the system's density matrix, $S_{\rm VN} = -\mathrm{Tr}\,(\rho_S\ln\rho_S)$. For a single Gaussian mode, this can be expressed in terms of the effective occupation number $n_k$:
\begin{equation}
    n_k = \frac{1-\sqrt{\gamma}}{2\sqrt{\gamma}} \, , \qquad
    S_{\rm VN} = 2\Big[(1+n_k)\ln(1+n_k) - n_k\ln n_k\Big] \, .
    \label{eq:ent_entropy}
\end{equation}
In this form, it is clear that $S_{\rm VN}$ grows monotonically as the system loses coherence ($\gamma \to 0$). Conversely, a monotonic decrease in entropy (increasing $\gamma$) signals partial \emph{recoherence}, a phenomenon known to occur in open de Sitter setups with very massive environmental field ~\cite{Colas:2022kfu}. Assuming an initial vacuum state requires $\gamma(\eta_0)=1$, implying $n_k(\eta_0)=0$ for both the system and environment at early times.

\begin{figure*}[t]
\centering
\begin{tabular}{cc}
\includegraphics[width=0.48\linewidth]{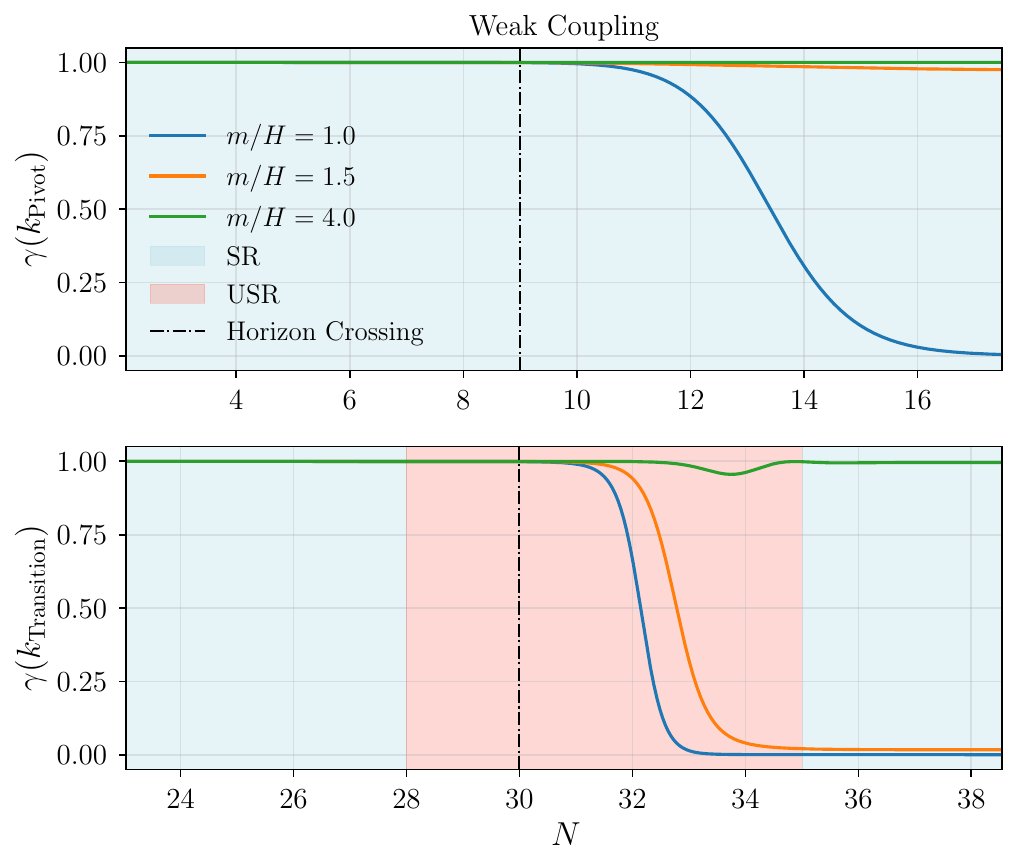} &
\includegraphics[width=0.48\linewidth]{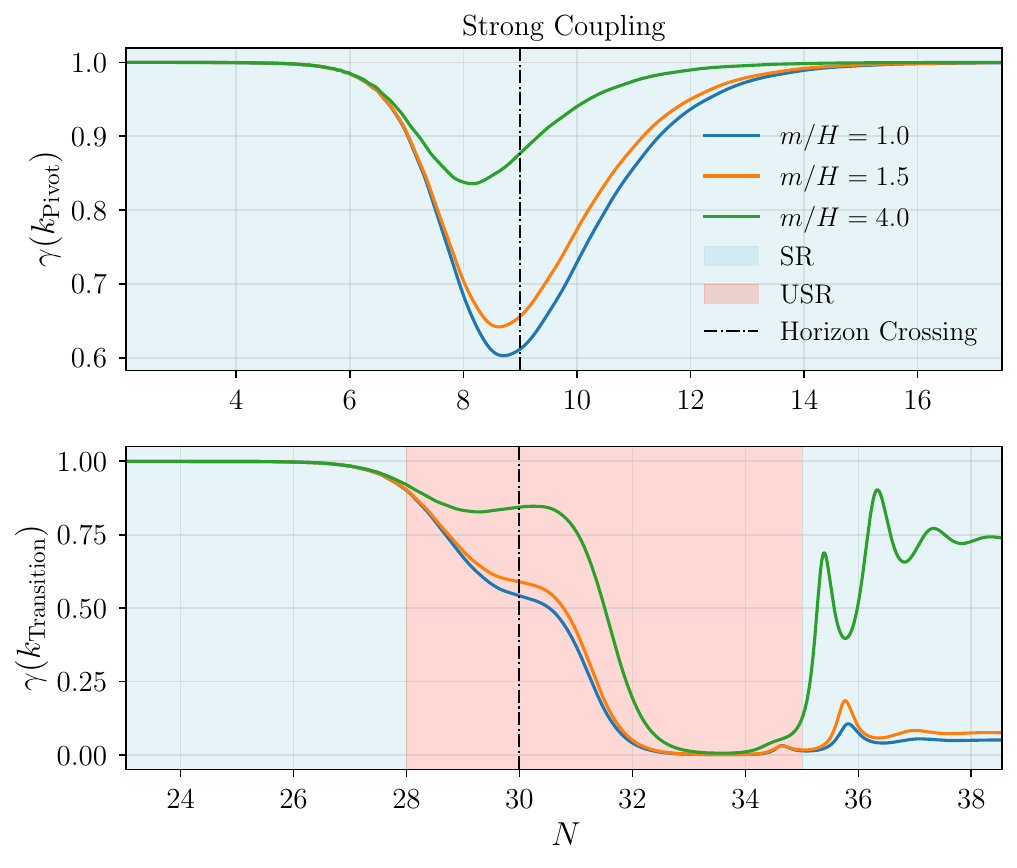} \\
\includegraphics[width=0.48\linewidth]{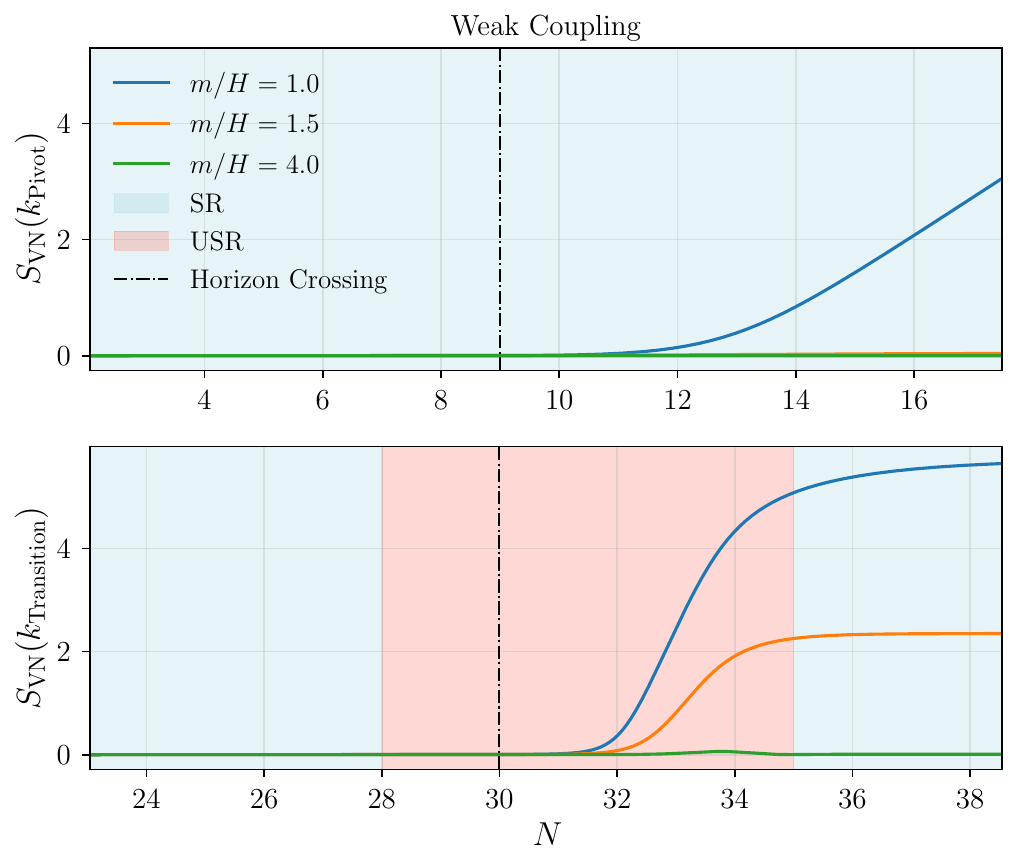} &
\includegraphics[width=0.48\linewidth]{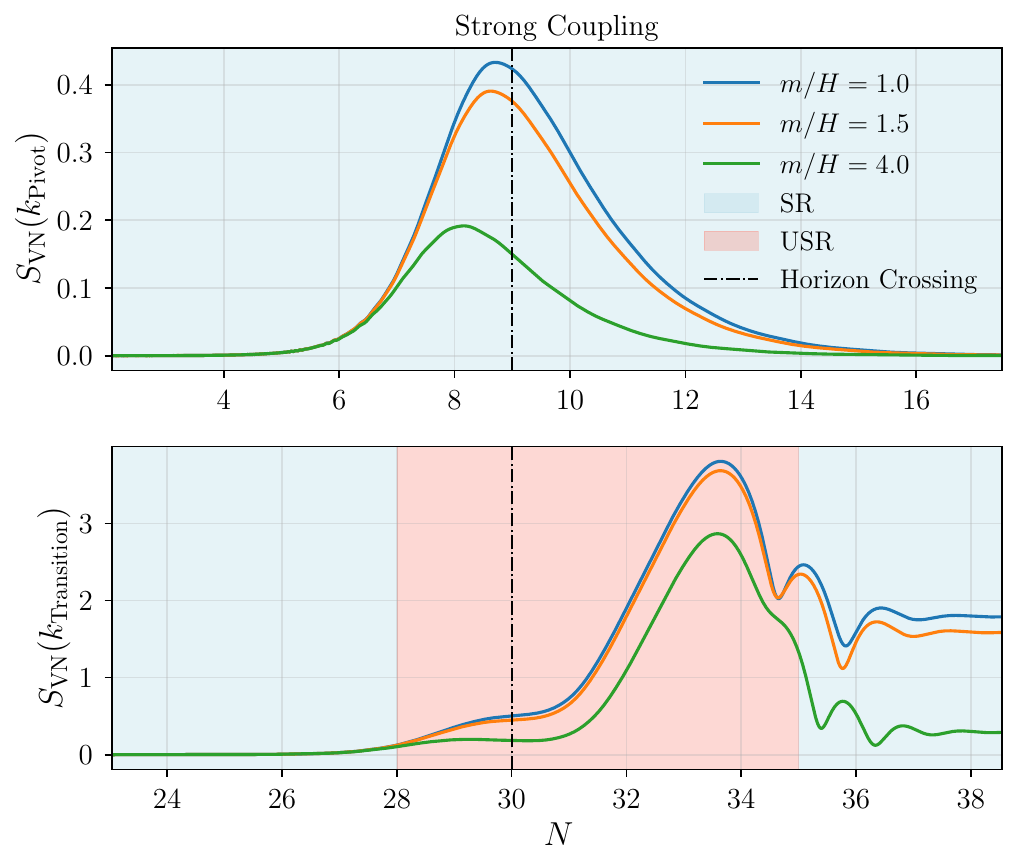}
\end{tabular}
\caption{Evolution of Purity $\gamma$ (top panels) and von Neumann entanglement entropy $S_{\rm VN}$ (bottom panels) under weak ($\lambda/H=0.05$, left) and strong ($\lambda/H=5.0$, right) environmental couplings. We track two representative modes: $k_{\rm Pivot} = 0.05 \, \mathrm{Mpc}^{-1}$, exiting the horizon deep in the initial SR phase, and $k_{\rm Transition} = 6.16 \cdot 10^7 \, \mathrm{Mpc}^{-1}$, exiting during the USR phase.}
\label{fig:purity_entropy}
\end{figure*}

\section{Results}
\label{sec:spectra}
\subsection{Decoherence Dynamics across the SR-USR-SR Transition}

Recently, the evolution of purity and entanglement entropy under non-attractor USR backgrounds has been investigated by \cite{Brahma:2024ycc}. Our full numerical solutions, derived via exact transport equations, are in striking agreement with their EFT findings, confirming that the efficiency of the quantum-to-classical transition is highly sensitive to the precise moment a mode crosses the horizon relative to the transient USR phase.

The resulting dynamics for purity and entropy are shown in Fig.~\ref{fig:purity_entropy}, for a first mode that exits the horizon during the SR phase ($k_{\rm Pivot}$) and a second mode crossing precisely during the USR regime ($k_{\rm Transition}$). The plots show that the mode evolution is in general strongly mass-dependent.

In the \textbf{weak coupling regime} ($\lambda/H=0.05$, left panels) and for the lightest environmental field ($m/H = 1.0$), the $k_{\rm Pivot}$ mode experiences monotonic, irreversible decoherence (and a corresponding rise in $S_{\rm VN}$). Such a decoherence is present also for the intermediate mass ($m/H = 1.5$), even if less evident from the figure, while for the heaviest mass ($m/H=4.0$) the system exhibits a distinctive non-monotonic trend: an initial decoherence is followed by an almost total \emph{recoherence} on super-Hubble scales, which restores the pure state of the system ($\gamma \simeq 1$, $S_{\rm VN} \simeq 0$). This behavior is perfectly consistent with standard SR benchmarks for heavy environments. The situation radically changes for modes exiting during the USR phase ($k_{\rm Transition}$). Here, the non-attractor kinematics overwhelmingly dominate the diffusive terms. Regardless of the environmental mass, the USR phase forces the mode to 
decohere violently and monotonically, with the sole exception of the 
heaviest configuration considered, $m/H=4$, which instead undergoes a 
partial recoherence process on super-Hubble scales, consistent with the 
standard decoupling of very massive environments already observed for 
$k_{\rm Pivot}$. Interestingly, as noted in Sec.~\ref{sec:spectra}, the most rapidly decoherence rate belongs to $m/H=1.0$, precisely the configuration that washes out the interference dip in the power spectrum (see next section).

In the \textbf{strong coupling regime} ($\lambda/H=5.0$, right panels), the interaction drastically modifies the dynamics of both modes. For $k_{\rm Pivot}$, the strong interaction induces early decoherence for all masses but then all of them, regardless of the environment mass, recover coherence at late times after the horizon crossing.  Conversely, the $k_{\rm Transition}$ mode undergoes severe decoherence exactly as it crosses the horizon during the USR phase. However, as the universe transitions back into the final SR phase, the strong environmental coupling forces a sharp, partial regain of purity. This is accompanied by ringing oscillations in both $\gamma$ and $S_{\rm VN}$ before stabilizing at an intermediate mixed state. 

These results clearly demonstrate that the USR phase is an incredibly efficient engine for generating entanglement and decoherence. However, the subsequent return to an SR attractor, especially under strong coupling, can dynamically alter this process, resulting in transient entropy decreases and partial recoherence that lock the quantum state into non-trivial configurations.

\subsection{Signatures of the Environment in the Power Spectrum}

\begin{figure*}[t]
\centering
\begin{tabular}{cc}
\includegraphics[width=0.48\linewidth]{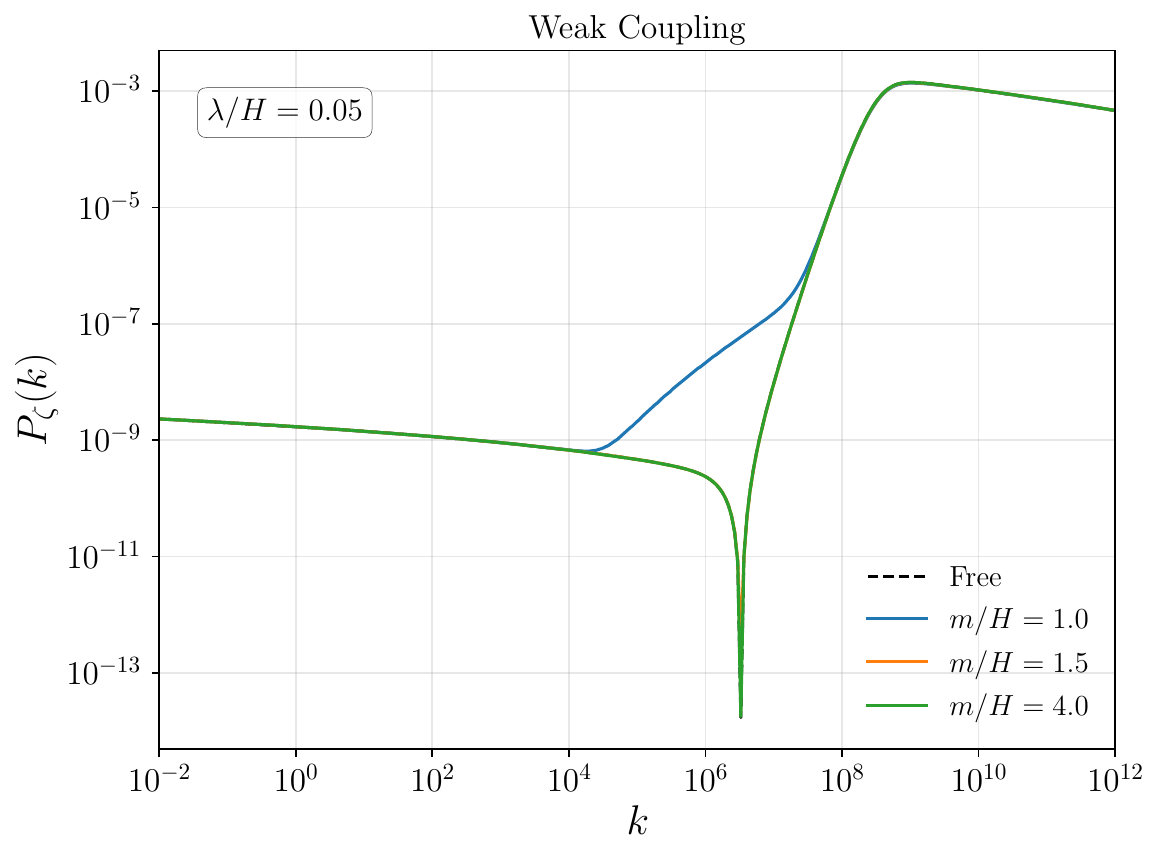} 
\hspace{3mm}
\includegraphics[width=0.48\linewidth]{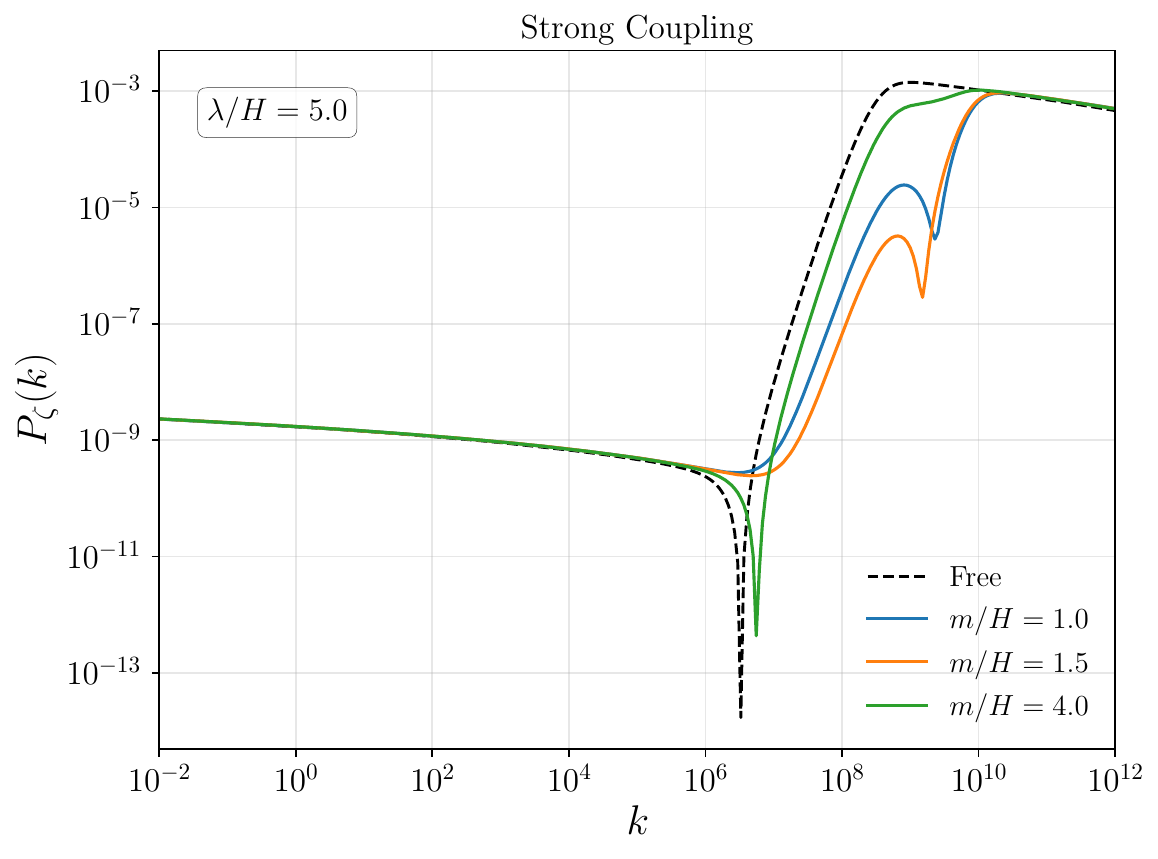} 
\end{tabular}
\caption{Primordial power spectrum of the curvature perturbation for weak ($\lambda/H=0.05$, left) and strong ($\lambda/H=5.0$, right) couplings to the entropic environment. The black dashed line represents the standard single-field (free) USR scenario. All spectra are normalized at the pivot scale, adopting the standard CMB normalization of $P_{\zeta} \sim 2.1 \times 10^{-9}$}.
\label{fig:powerspectrum}
\end{figure*}

From the definition of the primordial power spectrum of the adiabatic field $\zeta$, one can obtain an expression in terms of the reduced covariance matrix $\Sigma^{(\mathrm{S})}$:
\begin{equation}
P_{\zeta}(k) = \dfrac{\Sigma^{(\mathrm{S})}_{11}(k)}{z^2} \, ,
\end{equation}
where all quantities are evaluated in the super-horizon regime $k \ll aH$, and the factor $z^{-2}$ arises from the canonical rescaling $v = z\zeta$. In practice, we compute the super-horizon limit of $\Sigma^{(\mathrm{S})}_{11}$ by solving the transport equations for modes spanning a wide range of scales, $k \in [10^{-2}, 10^{12}] \, \mathrm{Mpc}^{-1}$. The resulting primordial power spectra are shown in Fig.~\ref{fig:powerspectrum}, illustrating the outcomes for different choices of the environmental mass $m$ and the interaction coupling $\lambda$.

A qualitative inspection of Fig.~\ref{fig:powerspectrum} reveals several striking features introduced by the coupling to the entropic sector. First and foremost, we observe that at very large ($k \lesssim 10^3 \, \mathrm{Mpc}^{-1}$) and very small ($k \gtrsim 10^{11} \, \mathrm{Mpc}^{-1}$) scales, the spectra asymptotically match the standard single-field predictions, represented by the black dashed line. These modes exit the Hubble radius deep within the first or the second SR phases, respectively, where the impact of the environment is efficiently decoupled. The crucial modifications occur strictly in the intermediate frequency window, where modes cross the horizon in the proximity of, or during, the USR phase. This specific window provides a unique observational opportunity to probe the physics of the entropic environment.

In the weak coupling regime ($\lambda/H=0.05$, left panel), the impact of the environment depends critically on the mass $m$. For environmental masses $m/H = 1.5$ and $m/H = 4$, the resulting power spectrum is practically indistinguishable from the standard single-field (free) case. However, a highly peculiar phenomenon emerges when the environmental mass exactly matches the Hubble scale, $m/H = 1$. In this specific scenario, the characteristic dip that typically precedes the steep power-law growth is completely washed out. Instead, the spectrum exhibits a smooth, approximately linear interpolation in log-scale connecting the initial slow-roll plateau directly to the rising slope, seamlessly continuing up to the main peak. Interestingly, this spectral feature strongly correlates with the quantum information dynamics discussed previously. As seen in the purity evolution for the weak coupling case, the lighter mass regimes exhibit total decoherence when crossing the horizon during the USR phase. Among these, the $m/H= 1$ case is precisely the one that drives the quantum-to-classical transition most rapidly, suggesting a deep link between the rapid loss of quantum coherence and the erasure of the destructive interference that generates the pre-growth dip. 
This coincidence admits a simple physical interpretation. The pre-growth dip 
in the free (single-field) USR spectrum originates from a destructive 
interference between the mode function's decaying and growing branches at 
horizon crossing. When the environmental mass satisfies $m/H\simeq1$, the 
entropic field becomes maximally efficient at exchanging phase information 
with the system precisely on the same time-scale set by the Hubble rate at 
which this interference builds up. The resulting resonant loss of phase 
coherence effectively randomizes the relative phase between the two branches 
before the interference pattern can develop, washing out the dip. For 
$m/H\gg1$ or $m/H\ll1$, the environment instead decouples adiabatically or 
too slowly to disrupt the interference within the transition window, leaving 
the dip intact.

The dynamics are drastically altered in the strong coupling regime ($\lambda/H=5.0$, right panel), where the interaction with the environment forces all mass cases to deviate significantly from the standard free scenario. Approaching the departure from the first SR phase, the pre-growth dip completely disappears for $m/H =  1$ and $m/H = 1.5$. Conversely, the $m/H = 4$ case still preserves a dip, but its position is visibly delayed, shifted to the right towards higher $k$ modes. 

Furthermore, during the transient phase leading to the USR enhancement, the strong environmental coupling induces a notable suppression of the maximum amplitude. None of the strongly coupled configurations reach the peak value of the free spectrum. Instead, they exhibit a complex oscillatory pattern: they rise to an intermediate peak, undergo a downward oscillation, and then bounce back, eventually reconnecting with the falling tail of the standard power spectrum at higher wave numbers. 

This phenomenon of suppressed and oscillatory peak structures could have profound implications for the physics of Primordial Black Holes (PBHs). Such modulations would directly alter the variance of the density perturbations, modifying the expected PBH mass fraction and formation thresholds. However, capturing the full extent of these implications, especially given the large amplitude of the perturbations, would ultimately require extending our study to include non-perturbative analyses, which we leave for future work.

\subsection{Scalar-Induced Gravitational Waves}

The substantial enhancements and distortions of the primordial power spectrum inevitably leave an imprint on the stochastic gravitational wave background. At second order in cosmological perturbation theory, the enhanced scalar fluctuations source tensor perturbations upon horizon reentry during the radiation-dominated era, producing a spectrum of scalar-induced gravitational waves (SIGWs) \cite{Ananda:2006af, Baumann:2007zm}. These second-order tensor perturbations could be much larger than those arising at first order from the free fluctuations of the metric tensor.

The dimensionless energy density of the induced GW background, $\Omega_{\rm GW}$, is generally computed by integrating the primordial scalar power spectrum over the relevant momentum scales \cite{Domenech:2021ztg, Ananda:2006af, Baumann:2007zm}:
\begin{equation}
\Omega_{\rm GW}(k) = 3 \int_0^\infty \!\! dv \int_{|1-v|}^{1+v} \!\! du \, \frac{T(u,v)}{u^2v^2} \mathcal{P}_{\mathcal{R}}(vk) \mathcal{P}_{\mathcal{R}}(uk) \, ,
\label{eq:OmegaGW_integral}
\end{equation}
where $u$ and $v$ are dimensionless integration variables representing the momenta of the sourcing scalar modes, and the integration kernel $\mathcal{T}(u,v)$ in a radiation-dominated universe is given by \cite{Domenech:2021ztg}:

\begin{equation}
\begin{split}
T(u,v) &= \frac{1}{4} \frac{1}{4v^2 - (1+v^2 - u^2)^2} \frac{4uv}{u^2 + v^2 - 3} \\
&\quad \times \Bigg[ \ln \frac{3 - (u+v)^2}{3 - (u-v)^2} - \frac{4uv}{u^2 + v^2 - 3} \\
&\quad + \pi^2 \Theta \left( u + v - \sqrt{3} \right) \Bigg] \, .
\end{split}
\label{eq:kernel_T}
\end{equation}

\begin{figure*}[t]
\centering
\begin{tabular}{cc}
\includegraphics[width=0.48\linewidth]{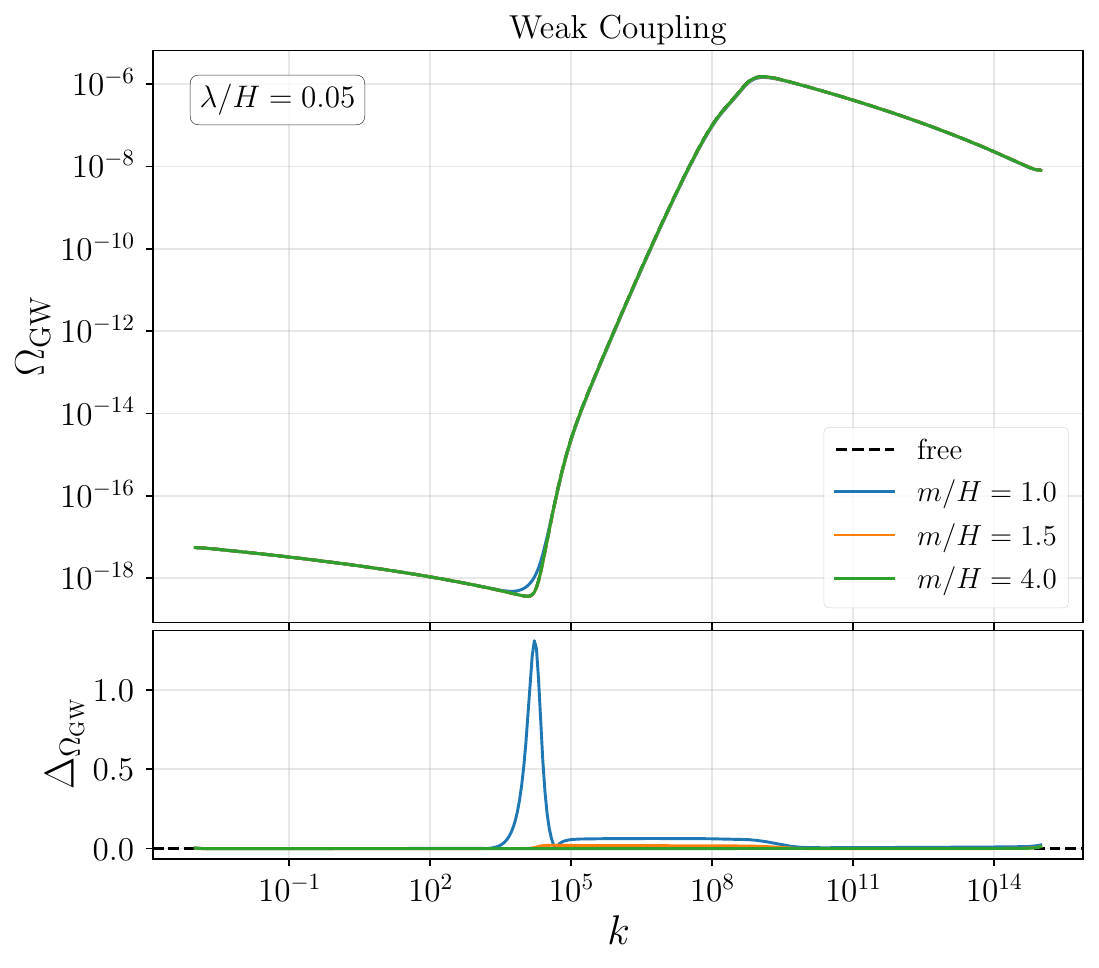} 
\hspace{5mm}
\includegraphics[width=0.48\linewidth]{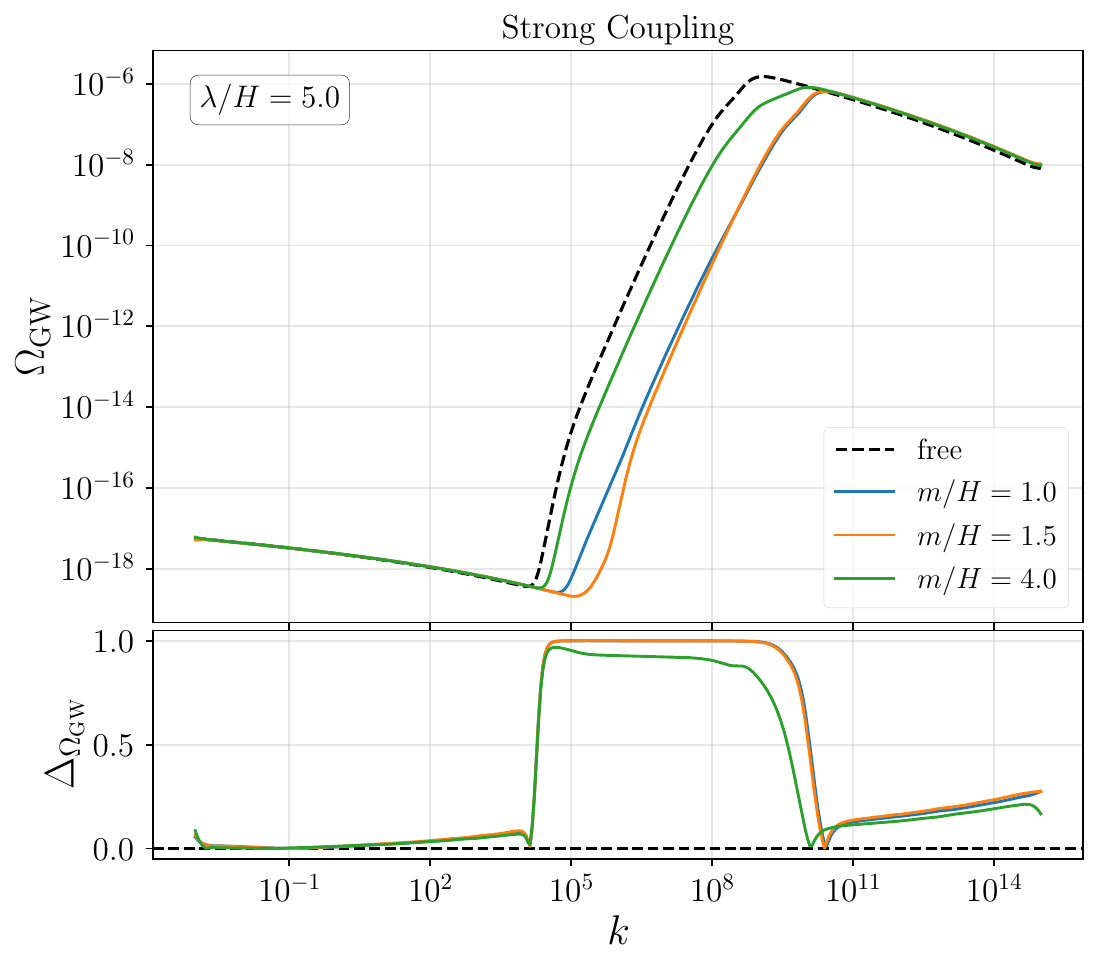}
\end{tabular}
\caption{Spectrum of Scalar-Induced Gravitational Waves $\Omega_{\rm GW}$ generated by the primordial scalar perturbations, evaluated for varying environmental mass $m$ at weak coupling $\lambda/H=0.05$ (left) and strong coupling $\lambda/H=5.0$ (right). The lower panels display the relative deviation $\Delta \Omega_{\rm GW}$ with respect to the standard single-field (free) case.}
\label{fig:IGWs}
\end{figure*}

We compute the induced gravitational wave spectral density, $\Omega_{\rm GW}$, using the Python package \textit{SIGWfast} \cite{Witkowski:2022mtg}, feeding our numerically exact primordial power spectra as the input source. The reconstructed induced gravitational wave signals for various environmental masses $m$ and couplings $\lambda$ are presented in Fig.~\ref{fig:IGWs}. 

Since the SIGW source term in Eq.~\eqref{eq:OmegaGW_integral} depends quadratically on $\mathcal{P}_\zeta$, the environmental features we identified in the scalar sector are further amplified and distinctly mapped into the GW signal. Analyzing the weak coupling regime ($\lambda/H = 0.05$, left panel), we observe that the GW spectrum is practically insensitive to the presence of the entropic environment. All mass variations yield an $\Omega_{\rm GW}$ profile that perfectly overlays the standard free USR scenario. The only visible exception, as shown in the relative difference plot ($\Delta \Omega_{\rm GW}$), is a highly localized spike at $k \sim 10^5 \, \mathrm{Mpc}^{-1}$ for $m/H=1$, which directly traces back to the erasure of the pre-growth dip in its corresponding scalar spectrum. 

Conversely, the strong coupling regime ($\lambda/H = 5.0$, right panel) displays dramatic, order-1 modifications. Because the scalar peaks in the strongly coupled scenarios are systematically suppressed compared to the free case, the resulting peak amplitudes of the GW spectra are correspondingly lowered by some orders of magnitude. The absence of the destructive interference dip in the scalar spectra for $m/H=1$ and $m/H=1.5$ translates into an altered, much broader infrared scaling of the GW peak. 

Furthermore, the right panel of Fig.~\ref{fig:IGWs} vividly shows that the strong environment shifts the maximum of the GW signal towards slightly higher frequencies. The oscillatory (ringing) features observed in the scalar spectrum also induce resonant modulations in the high-frequency tail of the GW profile.

These environment-induced deviations (such as the suppression of the peak, the altered infrared slopes, and the characteristic shifts) break the typical degeneracies of standard single-field USR models. Most importantly, these distinct signatures fall within a frequency range that could be probed by upcoming space-based interferometers like LISA, opening a potential new observational window to directly constrain the quantum nature of the inflationary environment and the physics of the early universe.

\section{Final Remarks and Conclusions}
\label{sec:conclusions}

The quantum-to-classical transition of primordial perturbations is not a 
mere formal curiosity: as we have shown, it leaves a concrete, calculable 
imprint on cosmological observables whenever inflation departs from a 
pure attractor trajectory. Using an effective two-field Lagrangian and the 
exact transport equations method, we tracked the full evolution of the system's covariance matrix across a transient USR phase, linking the microscopic loss of coherence between the adiabatic mode 
and its entropic environment to macroscopic features of the primordial 
power spectrum and its induced gravitational-wave signal.
Our analysis demonstrates that the interaction with the quantum environment produces key imprints on cosmological observables. Depending on the mass $m$ and the coupling $\lambda$ of the environment, the characteristic pre-growth interference dip typical of standard single-field USR can be completely erased, while the slope of the power spectrum can be significantly broadened. Furthermore, strong couplings induce marked oscillatory ringing features near the spectrum's peak. These environment-driven distortions in the primordial scalar spectrum quadratically propagate into the stochastic background of scalar-induced gravitational waves (SIGWs). We showed that strong environmental interactions lead to significant amplitude suppressions, altered infrared scalings, and resonant high-frequency modulations in the GW profile. These macroscopic deviations break single-field degeneracies and fall within frequency bands potentially accessible to future space-based interferometers like LISA.

Equally crucial is the timing of decoherence, as we found the efficiency of the quantum-to-classical transition to be exquisitely sensitive to the background kinematics at the time of horizon crossing. While modes exiting during standard slow-oll can exhibit partial recoherence (especially for heavy environmental masses), modes crossing the horizon during the non-attractor USR phase undergo rapid, violent, and irreversible decoherence, reaching a highly entangled state. Interestingly, the most rapidly decoherence rate, occurring for $m/H =  1$ at weak coupling, precisely correlates with the total erasure of the spectral dip, highlighting a profound link between the loss of phase coherence and structural modifications to the observable power spectrum.

While our findings shed new light on the role of quantum decoherence in non-attractor inflationary backgrounds, they also lay the groundwork for several compelling avenues of future research. A first natural extension concerns the inclusion of non-Gaussian correlators and the bispectrum transport. Because our current interaction model is purely Gaussian, the infinite hierarchy of the transport equations trivially closes at the level of the two-point correlation functions (the covariance matrix). However, PBH formation relies heavily on the extreme tail of the probability density function of curvature perturbations. Accurately modeling this requires evaluating non-Gaussian correlators. Extending our formalism to include cubic interactions would break the Gaussian closure, requiring us to dynamically couple the transport equations of the three-point correlators with the two-point system we solved here. Tracking the non-unitary evolution of the bispectrum through the USR phase would provide critical corrections to the PBH mass fraction when interactions with the environment are involved.

A second highly relevant direction involves exploring higher-spin environments and their connection to Cosmological Collider physics. In this study, we restricted the environment to a massive scalar field; however, string theory and supergravity embeddings of inflation generically predict a plethora of heavy fields with non-zero spin, such as vectors and tensors. Extending our effective Lagrangian to trace over higher-spin environments would introduce complex tensorial structures into the dissipation and diffusion kernels, potentially leaving unique angular signatures and distinct decoherence rates.

Ultimately, interpreting the primordial density fluctuations as an open quantum system provides a powerful lens through which to explore the early universe. As we move closer to the observational era of stochastic gravitational wave backgrounds, understanding the quantum mechanical origins of these signals will be paramount for disentangling the true physical nature of the inflationary epoch.

\begin{acknowledgments}
M.C. was supported by the European Research
Council through the ERC Synergy Grant COSMOMAG under grant No. 101224803. 
G.M.,  O.P. and S.S are supported by the Research grant TAsP
(Theoretical Astroparticle Physics) funded by Istituto Nazionale di Fisica Nucleare (INFN) and by Ministero dell’Università e della Ricerca (MUR), PRIN2022 program (Grant PANTHEON
2022E2J4RK) Italy.
\end{acknowledgments}

\appendix
\section{Transport Equations of Motion}
\label{app:sec:transpeq}

The transport equations governing the evolution of the covariance matrix are obtained 
by differentiating Eq.~\eqref{eq:covmatrix} with respect to conformal time in the 
Heisenberg picture and substituting the Heisenberg equations of motion for the 
phase-space operators $\hat{q}_i$. Since the total Hamiltonian is quadratic, this 
procedure yields a closed, exact system of first-order differential equations:
\begin{equation}
    \frac{d\Sigma}{d\tau} = \Omega \, H^{(\rm S+E)} \Sigma - \Sigma \, H^{(\rm S+E)} \Omega \, ,
\end{equation}
where $\Omega = \bigl(\begin{smallmatrix} 0 & 1 \\ -1 & 0 \end{smallmatrix}\bigr)$ 
is the symplectic matrix acting block-diagonally on the $4\times4$ phase space, 
and the full Hamiltonian matrix is explicitly given by
\begin{equation}
    H^{(\rm S+E)} = 
    \begin{pmatrix}
        k^2 & z'/z & 0 & 0 \\[4pt]
        z'/z & 1 & -\lambda a & 0 \\[4pt]
        0 & -\lambda a & k^2 + m^2 a^2 & a'/a \\[4pt]
        0 & 0 & a'/a & 1
    \end{pmatrix} \, .
\end{equation}

We adopt the number of e-folds $N$ as the time variable via $a(N) = a_0 \, e^N$, 
$z(N) \equiv \sqrt{2\epsilon(N)}\,a(N)$, and apply the chain rule 
$d/d\tau \to a(N)H(N)\,d/dN$. 
In this way, the ratio $z'/z = aH\big(1+\epsilon_2/2\big)$, which is an 
exact relation, linear in the slow-roll parameters, yet still 
fully valid in any inflationary background, attractor or not.
For compactness, we introduce the shorthand:
\begin{equation}
\theta_N \equiv \frac{d\ln z}{dN}\, , \qquad 
M^2 \equiv m^2+\lambda^2 \, ,
\end{equation}
so that the transport equations take the more compact form:
\vspace{-2.5em}
\begin{widetext}
\addtolength{\jot}{-4pt}
\begin{subequations}
\label{eq:transport-full}
\begin{align}
\intertext{\textit{System sector:}}
\frac{d\Sigma_{11}}{dN} &= 2\theta_N\,\Sigma_{11}+\frac{2}{aH}\Sigma_{12}-2\lambda\, \Sigma_{13}\, , \\[2pt]
\frac{d\Sigma_{12}}{dN} &= -\frac{k^2\Sigma_{11}-\Sigma_{22}}{aH}-\lambda\, \Sigma_{23}\, , \\[2pt]
\frac{d\Sigma_{22}}{dN} &= -2\theta_N\,\Sigma_{22}-\frac{2k^2}{aH}\Sigma_{12}\, ,
\intertext{\textit{Environment sector:}}
\frac{d\Sigma_{33}}{dN} &= 2\Sigma_{33}+\frac{2}{aH}\Sigma_{34}\, , \\[2pt]
\frac{d\Sigma_{34}}{dN} &= \frac{\lambda\, aH\,\Sigma_{23}-\big(k^2+M^2a^2H^2\big)\Sigma_{33}+\Sigma_{44}}{aH}\, , \\[2pt]
\frac{d\Sigma_{44}}{dN} &= 2\lambda\,\Sigma_{24}-\frac{2k^2}{aH}\Sigma_{34}-2M^2aH\,\Sigma_{34}-2\Sigma_{44}\, ,
\intertext{\textit{Mixing sector:}}
\frac{d\Sigma_{13}}{dN} &= (1+\theta_N)\,\Sigma_{13}+\frac{\Sigma_{14}+\Sigma_{23}}{aH}-\lambda\, \Sigma_{33}\, , \\[2pt]
\frac{d\Sigma_{14}}{dN} &= \lambda\,\Sigma_{12}+\theta_N\,\Sigma_{14}-M^2aH\,\Sigma_{13}-\Sigma_{14}-\frac{k^2\Sigma_{13}-\Sigma_{24}}{aH}-\lambda\,\Sigma_{34}\, , \\[2pt]
\frac{d\Sigma_{23}}{dN} &= \frac{-k^2\Sigma_{13}+(1-\theta_N)\,aH\,\Sigma_{23}+\Sigma_{24}}{aH}\, , \\[2pt]
\frac{d\Sigma_{24}}{dN} &= \lambda\,\Sigma_{22}-M^2aH\,\Sigma_{23}-\frac{k^2(\Sigma_{14}+\Sigma_{23})}{aH}-(1+\theta_N)\,\Sigma_{24}\, .
\end{align}
\end{subequations}
\end{widetext}
As discussed in Sec.~\ref{subsec:transpeq}, we supplement this system with an 
independent equation for the determinant of the reduced covariance matrix:
\begin{equation}
    \frac{d \det \Sigma^{(\rm S)}}{dN} = \Sigma_{11} \, \frac{d \Sigma_{22}}{dN} 
    + \Sigma_{22} \, \frac{d \Sigma_{11}}{dN} - 2\Sigma_{12} \, \frac{d \Sigma_{12}}{dN} \, ,
\end{equation}
yielding a coupled system of eleven differential equations solved numerically.

For the background, the initial scale factor $a_0$ is fixed by demanding that 
the pivot mode $k_{\rm Pivot} = 0.05 \, \mathrm{Mpc}^{-1}$ exits the Hubble 
radius $50$ e-folds before the end of inflation. For each mode $k$, Bunch-Davies 
initial conditions are imposed deep inside the Hubble radius at $k = 10^3\,aH$, 
and the evolution is tracked until the mode is well super-Hubble at 
$k = 10^{-5}\,aH$, which determines the numerical boundaries, $N_{\rm in}$ and 
$N_{\rm fin}$, respectively. All quantum information markers presented in this work 
are computed by evolving this system for a single mode $k$, while the primordial 
power spectrum is reconstructed by evaluating $\Sigma_{11}$ at super-Hubble scales 
over a broad range of modes $k \in [10^{-2},\, 10^{12}]\,\mathrm{Mpc}^{-1}$.

%\nocite{*}

\bibliography{apssamp}% Produces the bibliography via BibTeX.

\end{document}